\documentclass[usenatbib]{mn2e}
\usepackage{amsmath}
\usepackage{amsfonts}
\usepackage{amssymb}
\usepackage{graphicx}
\usepackage{pdfpages}
\begin{document}

\pagerange{\pageref{firstpage}--\pageref{lastpage}} \pubyear{2015}

\title[The Central Star Candidate of the Planetary Nebula Sh2-71]{The Central Star Candidate of the Planetary Nebula Sh2-71: Photometric and Spectroscopic Variability}

\author[T. Mo\v{c}nik, M. Lloyd, D. Pollacco and R. A. Street]{T. Mo\v{c}nik,$^{1,2}$\thanks{E-mail:
t.mocnik@keele.ac.uk} M. Lloyd,$^{3}$ D. Pollacco$^{4}$ and R. A. Street$^{5}$\\
$^{1}$Isaac Newton Group of Telescopes, Apartado de Correos 368, E-38700 Santa Cruz de La Palma, Spain\\
$^{2}$Astrophysics Group, Keele University, Staffordshire ST5 5BG, UK\\
$^{3}$Jodrell Bank Centre for Astrophysics, School of Physics and Astronomy, University of Manchester, Manchester M13 9PL, UK\\
$^{4}$Department of Physics, University of Warwick, Coventry CV4 7AL, UK\\
$^{5}$Las Cumbres Observatory Global Telescope Network, 6740 Cortona Drive, Suite 102, Goleta, CA 93117, USA}

\date{Accepted xxxx xxxxxxxx xx. Received xxxx xxxxxxxx xx; in original form xxxx xxxxxxxx xx}

\maketitle

\label{firstpage}

\begin{abstract}
We present the analysis of several newly obtained and archived photometric and spectroscopic datasets of the intriguing and yet poorly understood 13.5-mag central star candidate of the bipolar planetary nebula Sh2-71. Photometric observations confirmed the previously determined quasi-sinusoidal lightcurve with a period of 68 days and also indicated periodic sharp brightness dips, possibly eclipses, with a period of 17.2 days. In addition, the comparison between \textit{U} and \textit{V} lightcurves revealed that the 68-day brightness variations are accompanied by a variable reddening effect of $\Delta E(U-V)=0.38$. Spectroscopic datasets demonstrated pronounced variations in spectral profiles of Balmer, helium and singly ionised metal lines and indicated that these variations occur on a time-scale of a few days. The most accurate verification to date revealed that spectral variability is not correlated with the 68-day brightness variations. The mean radial velocity of the observed star was measured to be $\sim$26~km~s\textsuperscript{-1} with an amplitude of $\pm$40~km~s\textsuperscript{-1}. The spectral type was determined to be B8V through spectral comparison with synthetic and standard spectra. The newly proposed model for the central star candidate is a Be binary with a misaligned precessing disc.
\end{abstract}

\begin{keywords}
planetary nebula: individual: Sh2-71 -- stars: variables: general -- stars: emission-line, Be -- binaries: close.
\end{keywords}

\section{INTRODUCTION}

Planetary nebula (PN) Sh2-71 (PN G035.9-01.1, $\alpha=19^{\rmn{h}}02^{\rmn{m}}00.3^{\rmn{s}}$, $\delta=+2^\circ09^\prime11^{\prime\prime}$ J2000), see Fig.~1, was discovered by \citet{Minkowski46} who classified it as a diffuse and peculiar nebulosity. In 1959 the second Sharpless catalogue of emission nebulae listed the object as a possible PN \citep{Sharpless59}. Modern catalogues classify the object as a true PN \citep{Kohoutek01}.

\begin{figure*}
\fbox{\includegraphics[width=12cm]{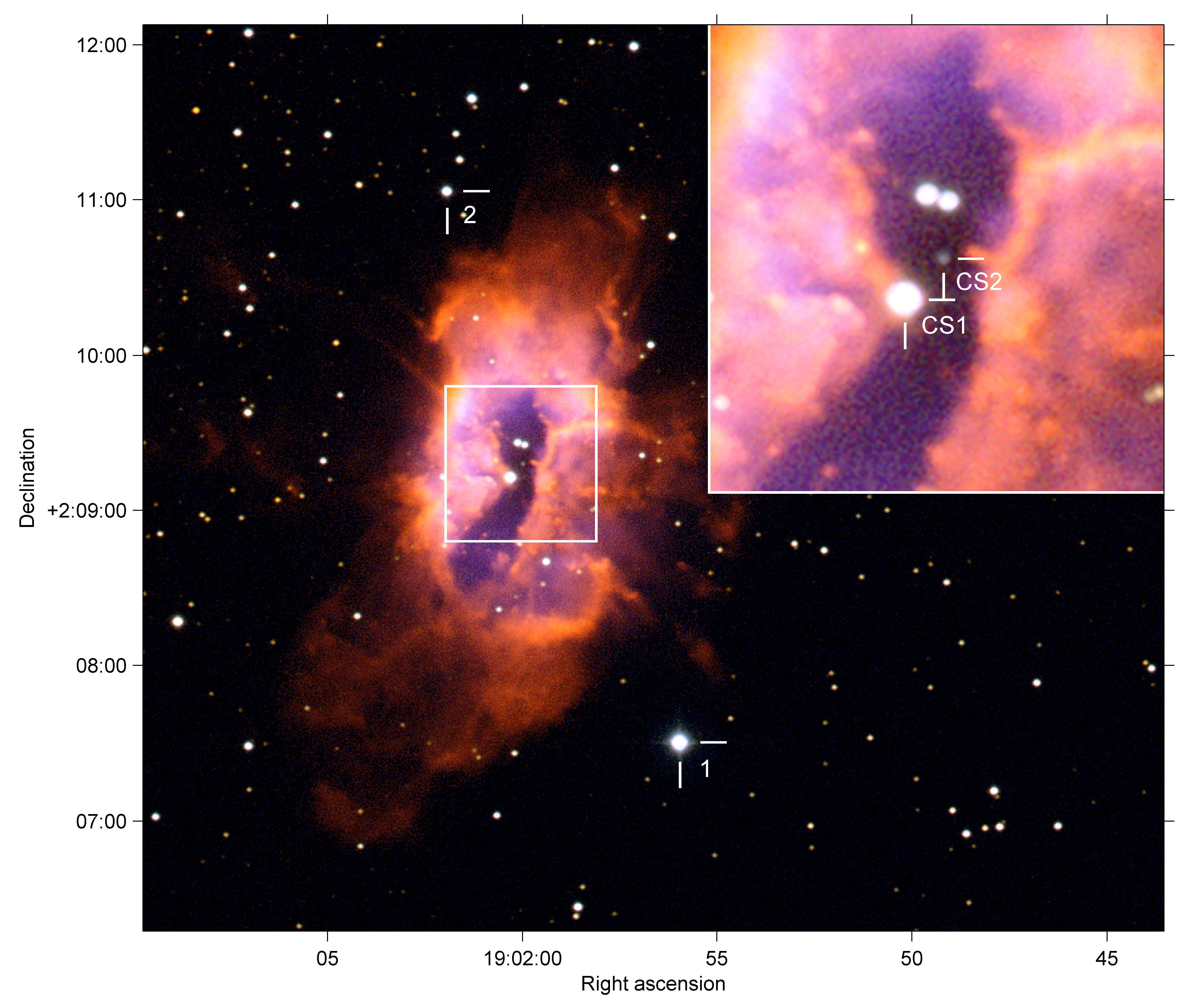}}
\caption{RGB image of planetary nebula Sh2-71 and zoomed central region of the nebula (top right panel), identifying central star candidates CS1 and CS2.  Note the nebular feature which is extending eastwards from CS1. The two comparison stars, which were used for the differential photometry, are labelled 1 and 2. H$\alpha$ (red), Sloan r' (green), [O\thinspace\textsc{iii}] (blue).}
\end{figure*}

The first extensive photometric observations of the variable 13.5-mag apparent central star of Sh2-71 were performed by \citeauthor{Kohoutek79} between 1977 and 1979. \citet{Kohoutek79} reported brightness variability of more than 0.7~mag with indications of a smooth sine-like lightcurve. He also used the star's colour indices to initially classify the star as a B8V and noted that the spectral type B8V cannot be responsible for the observed high excitation in the surrounding PN. Therefore, he suggested that the central star is a binary system in which the secondary star is hot enough to ionize the nebula. The largest photometric study to date was presented by \citet{Mikulasek07}. They obtained 2,004 photometric measurements in \textit{V}, \textit{R}$_\rmn{C}$ and \textit{I}$_\rmn{C}$ band, and combined them with existing photometric data from \citet{Kohoutek79} and \citet{Jurcsik93}. From the combined 4,268 measurements, spanning over 28 years, they identified the brightness variations to be a quasi-sinusoidal lightcurve with a period of $68.101 \pm 0.010$ days, and that the lightcurve minima flatten occasionally. \citet{Mikulasek07} also noted pronounced spectral variations between two spectra, taken at different brightness phases.

A detailed model for the central star as the progenitor of the collimated bipolar nebula was suggested by \citet{Cuesta93}. Primarily based on their four slit spectra covering both star and nebula and taken on two consecutive nights they suggested that the central star is a close binary system in which the material is being transferred from the main-sequence star to an ionising star. They further speculated that the material expelled through the second Lagrangian point creates the circumstellar disc, which collimates the fast stellar wind into a bipolar shape. Based on synthetic composite spectra, \citet{Cuesta93} determined the spectral type of the cooler, brighter component to F0V-A7V and estimated the effective temperature of the ionising star to be 10\textsuperscript{5}~K. To date, there is no direct evidence for a hot, ionising binary component.

Although the variable 13.5-mag star has historically been identified as the central star (star CS1 in Fig.~1), \citet{Frew07}, have suggested however that a nearby, much dimmer and bluer, 19-mag star may be the true central star (star CS2 in Fig.~1). \citet{DeMarco08} have concluded that there is insufficient evidence either way to give a definitive identification of the true central star.

In this paper we present our imaging of the nebula and challenge the alternative suggestion for the central star. Furthermore, we present our analysis of new photometric and spectroscopic observations, coupled with some archived and yet unpublished spectroscopic datasets. From these observations we derive several new parameters of the brighter central star candidate and refine certain previously determined properties. We suggest a new model for the central star candidate in order to establish a link between the observed properties and the collimated bipolar shape of the PN.

\section{OBSERVATIONS AND DATA REDUCTION}

We have imaged Sh2-71 in broad- (Sloan r', 3$\times$30~s) and two different narrowband filters (H$\alpha$ and [O\thinspace\textsc{iii}], both 3$\times$120~s) on 2013 October 21 with the Wide Field Camera (WFC), which is mounted on the 2.5-m Isaac Newton Telescope (INT) on La Palma in the Canary Islands. The images were processed in the usual way using \textsc{theli}, \textsc{fits liberator} and \textsc{matlab} to produce the RGB image shown in Fig.~1.

The three photometric datasets of the $\sim$13.5-mag central star candidate CS1, presented in this paper, were obtained with the 0.8-m telescope at Byrne Observatory at Sedgwick (BOS) in California, the 2-m Faulkes Telescope South (FTS) in Siding Spring, Australia, and FTS's twin telescope, the 2-m Liverpool Telescope (LT) on La Palma. Observational details are listed in Table~1. All three photometric datasets have been bias subtracted and flat field corrected by the automatic pipelines, provided by the individual observatories. Combining the LT images and differential aperture photometry were carried out using the \textsc{starlink} application packages \textsc{ccdpack} and \textsc{photom}. The non-variable comparison star, which was present in every image and yielded the highest signal to noise ratios of the measured differential magnitudes for CS1 in FTS and LT datasets, is shown in Fig.~1 as comparison star number 1. Because the BOS images were obtained with much longer exposure times compared to FTS and LT, the comparison star number 1 lay outside the linear range of the CCD's full well capacity for some of the BOS images. Thus, the BOS differential photometry was measured with a fainter comparison star which is labelled as comparison star number 2 in Fig.~1. The differential photometry was performed using circular photometric apertures and annulus sky subtraction. The aperture radii were set to be equal to one average stellar full width at half maximum (FWHM) as measured for a set of five most suitable stars in each image. Sky annulus was placed between two and three FWHM from the stellar centres. This photometric configuration was found to maximize the signal to noise ratios of the differential magnitudes. We were only able to measure the differential photometry for CS1; CS2 was too faint to be detectable in the FTS and LT datasets and below the 5-sigma detection limit for the majority of the images in the BOS dataset. The log of the final differential magnitudes for CS1 is given in Online Appendix in Tables~1.1-1.5.

\begin{table*}
\centering
\begin{minipage}{110mm}
\caption{Observational details for photometry. The table lists the names of the telescopes, the dates of the first and last observation, applied filters, exposure times, mean signal to noise ratios for the $\sim$13.5-magnitude CS1 and the final number of useful photometric data points.}
\begin{tabular}{@{}cccccc@{}}
\hline
Telescope&Duration of &Filter&Exposure&S/N&Number of\\
         &observations&      &time (s)&   &data points\\
\hline
BOS&2011 Jun 26 $-$ 2011 Jul 28&\textit{B}&600&543&21\\
FTS&2011 Oct 4 $-$ 2012 Mar 23&\textit{B}&15&92&7\\
&&\textit{V}&15&157&7\\
LT &2012 Mar 31 $-$ 2012 Jul 20&\textit{U}&3$\times$24&64&34\\
&&\textit{V}&3$\times$3&171&36\\
\hline
\end{tabular}
\end{minipage}
\end{table*}

The first spectroscopic dataset comprises 25 unpublished, archived slit spectra of the central star candidate CS1 which were obtained from the Intermediate Dispersion Spectrograph (IDS), mounted on the INT. A further 9 archived and unpublished high-cadence and high-resolution echelle spectra were obtained from the University College London Echelle Spectrograph (UCLES) on 3.9-m Anglo-Australian Telescope (AAT) in Siding Spring, Australia. The third spectroscopic dataset, purposely obtained for this research project, was obtained from the Fibre-Fed Robotic Dual-Beam Optical Spectrograph (FRODOSpec) at the LT. Observational details for the individual spectroscopic datasets are presented in Table~2. Pre-processing, extraction of spectra and wavelength calibration of the INT dataset were performed with \textsc{starlink} application packages \textsc{figaro} and \textsc{echomop}. The AAT spectra were reduced with the \textsc{iraf} application package \textsc{echelle}. The LT spectra were reduced with the fully automated data reduction pipeline provided by \citet{Barnsley12}.

\begin{table*}
\centering
\begin{minipage}{145mm}
\caption{Overview of spectroscopic observations. The INT exposure time is given as a median value. Mean signal to noise ratios are calculated for central wavelengths, except for AAT, where it is provided for H$\beta$ region.}
\begin{tabular}{@{}ccccccc@{}}
\hline
Telescope&Duration of &Wavelength &Resolving &Exposure&S/N&Number of\\
         &observations&range (\AA)&power&time (s)&   &spectra\\
\hline
INT (IDS)&1996 Jul 10 $-$ 1999 May 4&3820$-$4800&4500&1500&161&25\\
AAT (UCLES)&2000 May 21&3700$-$5100&44000&1800&20&9\\
LT (FRODOSpec)&2012 Jun 16 $-$ 2012 Jul 19&3900$-$5100&5500&1200&16&2\\
&&5900$-$8000&5300&1200&31&3\\
\hline
\end{tabular}
\end{minipage}
\end{table*}

\section{RESULTS}

\subsection{Imaging}

Our RGB image of Sh2-71 is shown in Fig.~1. As noted in the Introduction, the bright star CS1, historically believed to be the central star of Sh2-71, is not hot enough to ionise the nebula. This led \citet{Frew07} to propose that the faint ($19^{\rm th}$ mag) blue star, offset 7~arcsec NW of CS1 and identified here as CS2, may be the true central star. They claim that the measured brightness of CS2 corresponds to the expected luminosity of the progenitor's remnant at the nebula's distance of 1~kpc. In addition, \citet{Frew07} note that CS2 is also closer to the apparent geometric centre of the nebula, whereas CS1 is slightly offset from the centre. The WFC images of the nebula have been used to investigate the nebular features close to the centre of the nebula to try to find evidence as to which star is the true central star of the PN.

The central position of CS2 favours its identification as the central star, although several other PNe have been discovered to possess off-centre central stars \citep{DeMarco09}. However, the somewhat symmetrical nebular structure about a line eastwards from CS1 may suggest a physical link to this star. Furthermore, since the most likely collimating mechanism for the bipolar shape of a PN is a binary central star \citep{DeMarco09}, the progenitor star of the bipolar PN Sh2-71 is expected to have a companion star which would provide additional brightness contribution. Thus, a brighter central star candidate, CS1, seems a more plausible choice for a binary central star hypothesis. In such a case, CS1 would comprise a hot ionising component and a cooler, optically brighter companion which would dominate the observed optical flux, effectively hiding the hot ionising companion. A similar binary system has been identified as the central stellar system of another bipolar PN, NGC~2346 (\citet{Mendez81}, also see subsection 4.2). Finally, the reported mean RV of the PN and measured mean stellar RV of CS1 are very similar (see subsection 3.3.3), as would be expected for a central star and its planetary nebula, although as yet there is no RV measurements for CS2. Thus CS1 remains a reasonable candidate for the central star at this stage.

\subsection{Photometry}

The complete detailed photometric analysis is presented in \citet{Mocnik13}. In this paper the main photometric results are summarised in Fig.~2, which shows a set of four individual lightcurves obtained from three different telescopes. The BOS \textit{B}, FTS \textit{V} and LT \textit{V} datasets are used to reveal the shape of the lightcurve, whereas the overlay between LT \textit{V} and LT \textit{U} illustrates the difference in axes scaling between differential magnitudes $\Delta V$ and $\Delta U$, which can be used to estimate the reddening effect. Inconsistencies in the filters used and the available comparison stars for each image make it difficult to compare these photometric measurements directly. Here, the BOS data have been obtained using comparison star 2 (see Fig.~1) and the FTS and LT data have been obtained using comparison star 1. Thus the FTS and LT datasets are directly comparable, however, the BOS differential magnitudes had to be offset. Fig.~2 also shows the predicted lightcurve given by \citet{Mikulasek07}.

\begin{figure*}
\fbox{\includegraphics[width=\textwidth]{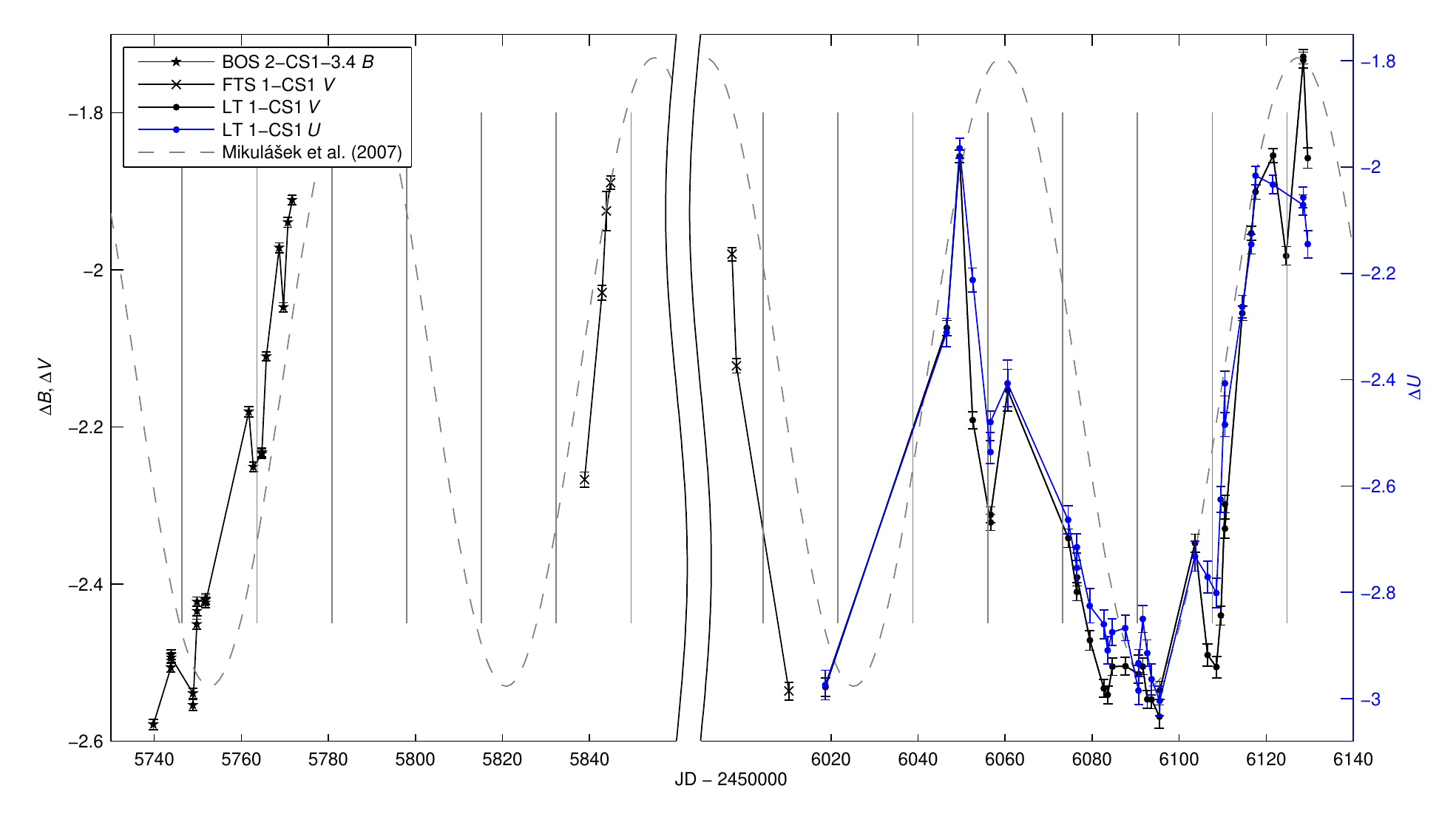}}
\caption{The comparison between predicted and measured lightcurves of CS1 with tentative indication of sharp periodic brightness dips. For compatibility reasons, the BOS differential magnitudes were offset by -3.4~mag. The dashed line represents the predicted lightcurve by \citet{Mikulasek07}. Potentially periodic brightness dips are indicated by engraved vertical lines with a period of 17.2 days. The difference in axes scaling between LT \textit{V} (black) and \textit{U} (blue) lightcurves suggests a variable reddening effect. The data points are connected with lines for easier presentation purposes.}
\end{figure*}

First of all, we can see from Fig.~2 that the combined lightcurve is in rough agreement with the period and phase reported by \citet{Mikulasek07} (JD$_{\rm max} = 2\,449\,862.02 \pm 0.32 + 68.101 \pm 0.010(E-96)$). The comparison confirms that the photometric nature of CS1 has remained the same since the last Mikul\'{a}\v{s}ek's observation in 2002. The disagreement between the observed and predicted decreasing part of the lightcurve is most likely due to the current deviation of the lightcuve's shape from the assumed sinusoidal brightness variations.

Secondly, several sharp dips can be seen in the lightcurve. These are significantly larger than the uncertainties of individual data points but much smaller than the almost 1-mag peak-to-peak variation in the overall lightcurve. Attempts at a formal periodic analysis \citep{Mocnik13} were unsuccessful due to insufficient data. However, a simple visual fit of these dips, made by trying various periods and epochs, revealed that the five most pronounced brightness dips can be characterised as periodic with a period of 17.2$\pm$0.1 days. The overlap between the brightness dips and best-fit epoch and period is illustrated with the periodic vertical lines in Fig.~2. Such a potentially periodic behaviour points to the intriguing possibility of an eclipsing binary system, however, many more data points would be required for a more reliable and accurate analysis of these possibly periodic brightness dips.

Lastly, the difference between the axes scales for LT \textit{V} and \textit{U} filter in Fig.~2 reveals that the brightness variation amplitude is larger for \textit{U} filter and therefore that the central star exhibits a variable reddening effect along with the brightness variations. The variable component of reddening effect was estimated to be $\Delta E(U-V)=0.38$. On the other hand, $\Delta E(B-V)$ could not be estimated reliably because BOS \textit{B} lightcurve did not chronologically overlap with any other photometric dataset, and because of insufficient number of data points in FTS \textit{B} dataset.

\subsection{Spectroscopy}

\subsubsection{Spectroscopic Variability}

Fig.~3 shows two fully reduced and normalized spectra of the observed star near the H$\beta$ line, obtained with the LT at different phases of the 68-day lightcurve. Pronounced spectral variations of the emission components of Balmer and singly ionised metal line profiles confirm the variations, reported by \citet{Mikulasek07}. However, in contrast to Mikul\'{a}\v{s}ek's reporting we did not notice pronounced [O\thinspace\textsc{iii}] lines in any of the retrieved spectral datasets. Possibly the spectra analysed by \citeauthor{Mikulasek07} were contaminated by the nebular [O\thinspace\textsc{iii}] line.

\begin{figure}
\fbox{\includegraphics[width=8cm]{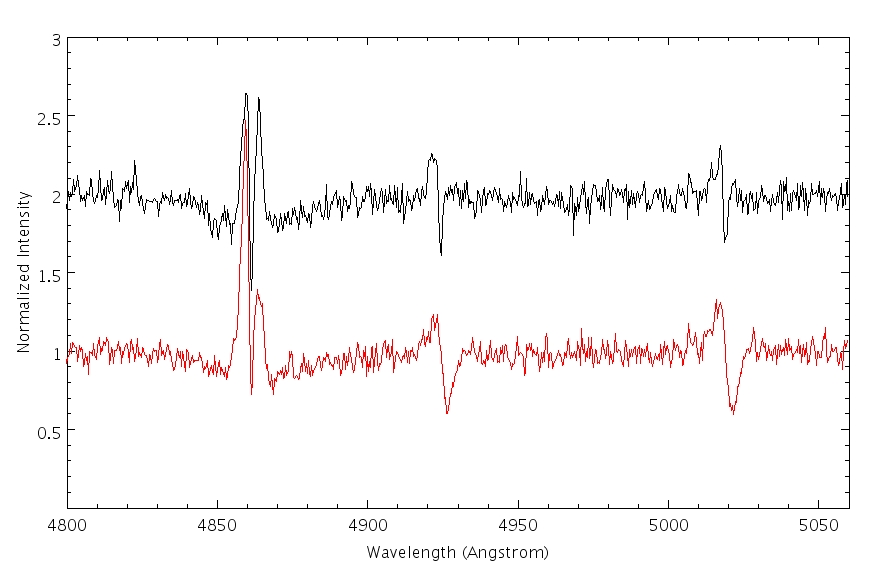}}
\caption{LT spectra of CS1 near the H$\beta$ profile. The two spectra demonstrate the pronounced variations of H$\beta$ (4861~\AA) and Fe\thinspace\textsc{ii} lines (4924 and 5018~\AA). The first spectrum, plotted with the black colour, has been obtained on 2012 July 1 with a corresponding 68-day brightness phase of 0.754. The second spectrum, shown in red, was obtained on 2012 July 19 with a phase of 0.018.}
\end{figure}

We employed a dataset of 25 archived INT slit spectra to provide the most accurate verification to date if the spectral variations are correlated to 68-days brightness variations. For each of the reduced and normalized spectrum we have calculated the corresponding brightness phase from the ephemeris, published by \citet{Mikulasek07}. Normalized INT spectra, arranged according to their calculated brightness phase, are shown in Fig.~4.

\begin{figure*}
\fbox{\includegraphics[width=\textwidth]{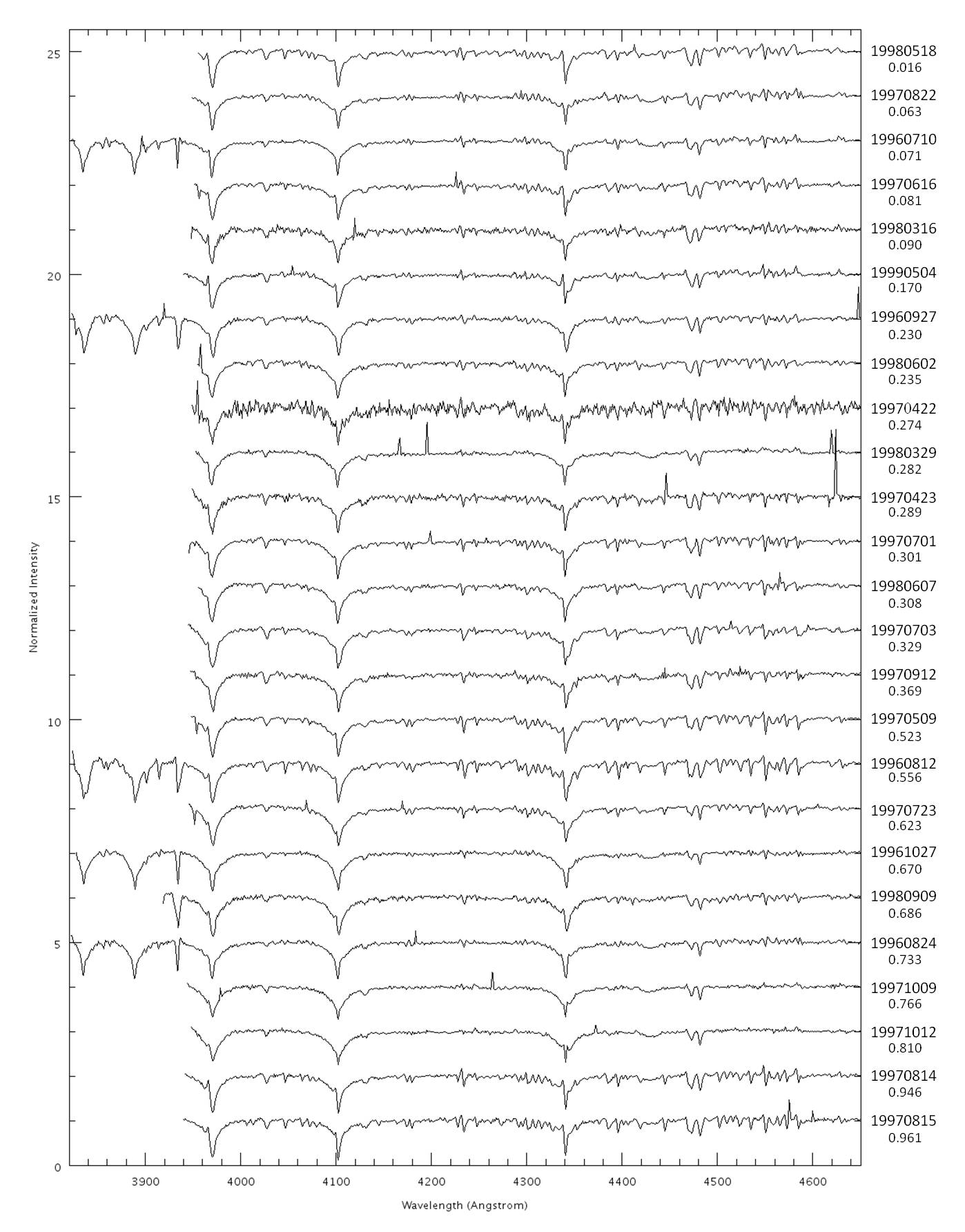}}
\caption{A set of INT spectra, ordered by their phase within the 68-day photometric variability. The phases of individual observing nights are provided under each corresponding date. The most pronounced spectral variability may be noticed for Balmer series, singly ionised metal lines and neutral helium (4026 and 4471~\AA). Note that the spectral variability is not correlated with the brightness phase.}
\end{figure*}

In contrary to expressed expectations in the previous papers related to Sh2-71, Fig.~4 demonstrates that the spectral variations are not correlated with 68-days brightness variations. Due to a large chronological spread of the archived INT spectra we were unable to determine any other potential period in the spectral variability. For this task the spectra would have to be obtained with a higher cadence, preferably with a daily cadence. On the other hand, 9 high-resolution echelle spectra from the AAT with a cadence of 30 minutes did not reveal any noticeable spectral variations in the time-span of 4.5 hours (see Fig.~5). The presence of the spectral variations in the LT and INT datasets and their absence in the AAT dataset imply that the spectral variations must occur on a time-scale of a few days.

\begin{figure}
\fbox{\includegraphics[width=8cm]{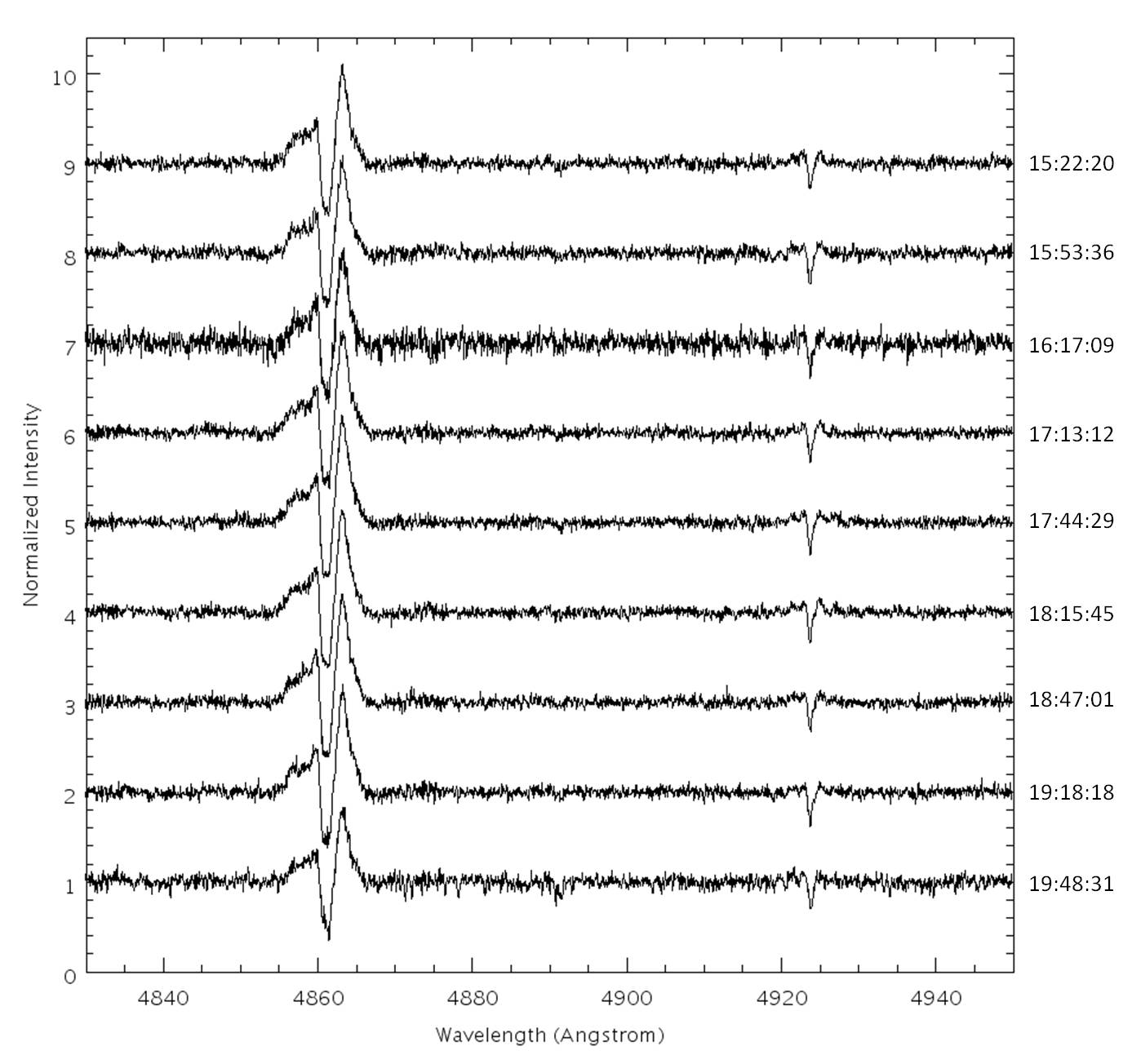}}
\caption{AAT spectra of CS1 near the H$\beta$ (4861~\AA) and Fe\thinspace\textsc{ii} (4924~\AA) lines. Shown on the right are the mid-exposure UTs. Note the absence of pronounced spectroscopic variations in the 4.5 hour time-span and that H$\beta$ emission components are reversed compared to the LT spectra in Fig.~3.}
\end{figure}

\subsubsection{Spectral Type}

We compared the measured spectra of the central star candidate CS1 with synthetic spectra, computed by \citet{Munari05}. The best-fitted stellar atmospheric parameters were determined to be \textit{T}$_\rmn{eff}$ 11000--13000~K, $\log(g)$ $4.0\pm0.25$~cm~s\textsuperscript{-2}, $v_\rmn{rot}\cdot \sin(i)$ $200_{-25}^{+50}$~km~s\textsuperscript{-1} with an indicated high value of metallicity. These parameters classify the observed star as a B9--B7 main-sequence star. The further comparison to standard spectra, provided by \citet{Bagnulo03}, confirmed a late B main-sequence spectral type. The comparisons and details regarding the synthetic and standard spectra are presented in Online Appendix and \citet{Mocnik13}.

Our $\sim$B8V spectral type classification of CS1 is in agreement with earlier B8V classification by \citet{Kohoutek79}, who determined the spectral type by colour indices. On the other hand, our $\sim$B8V classification refutes the spectral type A7V-F0V  by \citet{Cuesta93} who were using slit spectra in the wavelength range 6000--7400~\AA. They noted, however, the presence of Si\thinspace\textsc{ii} lines (6347 and 6371~\AA) in their spectra, which correspond to stars with temperatures higher than A7V. The spectral wavelength range, covered by this project's INT spectra, has revealed many more spectral features that cannot be explained with an A7V star, such as the presence of Si\thinspace\textsc{ii} lines (3856 and 3863~\AA), He\thinspace\textsc{i} (4026 and 4471~\AA) and absence of Mn\thinspace\textsc{i} (4033~\AA), Ca\thinspace\textsc{i} (4227~\AA) and Ti\thinspace\textsc{ii} (4468~\AA). The synthetic spectra fitting and comparison with the standard spectra provide strong evidence that the spectral type of the central star candidate CS1 is $\sim$B8V.

\subsubsection{Radial Velocity}

Radial velocities (RVs) of CS1 in the INT and AAT datasets have been measured in two steps. First, we measured the relative RVs with respect to the first spectrum in each of the two datasets. For the INT dataset we used the \textsc{starlink} tool \textsc{hcross}, the template INT spectrum of CS1 from 1996 Jul 10 and a cross-correlation wavelength range 3960--4600~\AA. For the AAT dataset we used the \textsc{iraf} tool \textsc{fxcor}, template AAT spectrum of CS1 with mid-exposure time 15:22:20 UT and a cross-correlation wavelength range 4000--5000~\AA. Additional binning was not applied to either of the two datasets. The second step was to convert relative RVs into absolute values. This was done for both datasets by measuring the wavelength of Mg\thinspace\textsc{ii} 4481~\AA, which has been identified as the most stable and suitable individual spectral line for measuring RVs in our spectra. The final absolute heliocentric RVs for the INT and AAT datasets are listed in Online Appendix in Tables 2.1 and 2.2, respectively.

Our INT RV measurements revealed that the mean RV of the central star candidate CS1 is $\sim$26~km~s\textsuperscript{-1} and that RV varies with an amplitude of $\pm$40~km~s\textsuperscript{-1} (see top panel in Fig.~6). Due to the large spread in observation time of our INT spectra, we were unable to determine the period of a potential periodicity of the RV variations (see bottom panel in Fig.~6). However, the cadence and the number of data points were sufficient to reject any correlation with the 68-days brightness variations (see middle panel in Fig.~6). In addition, because of the absence of RV variations in AAT spectra with 30-minutes cadence (see Table 2.2 in Online Appendix) we also rejected hourly time-scales for a potential RV periodicity.

\begin{figure}
\fbox{\includegraphics[width=8cm]{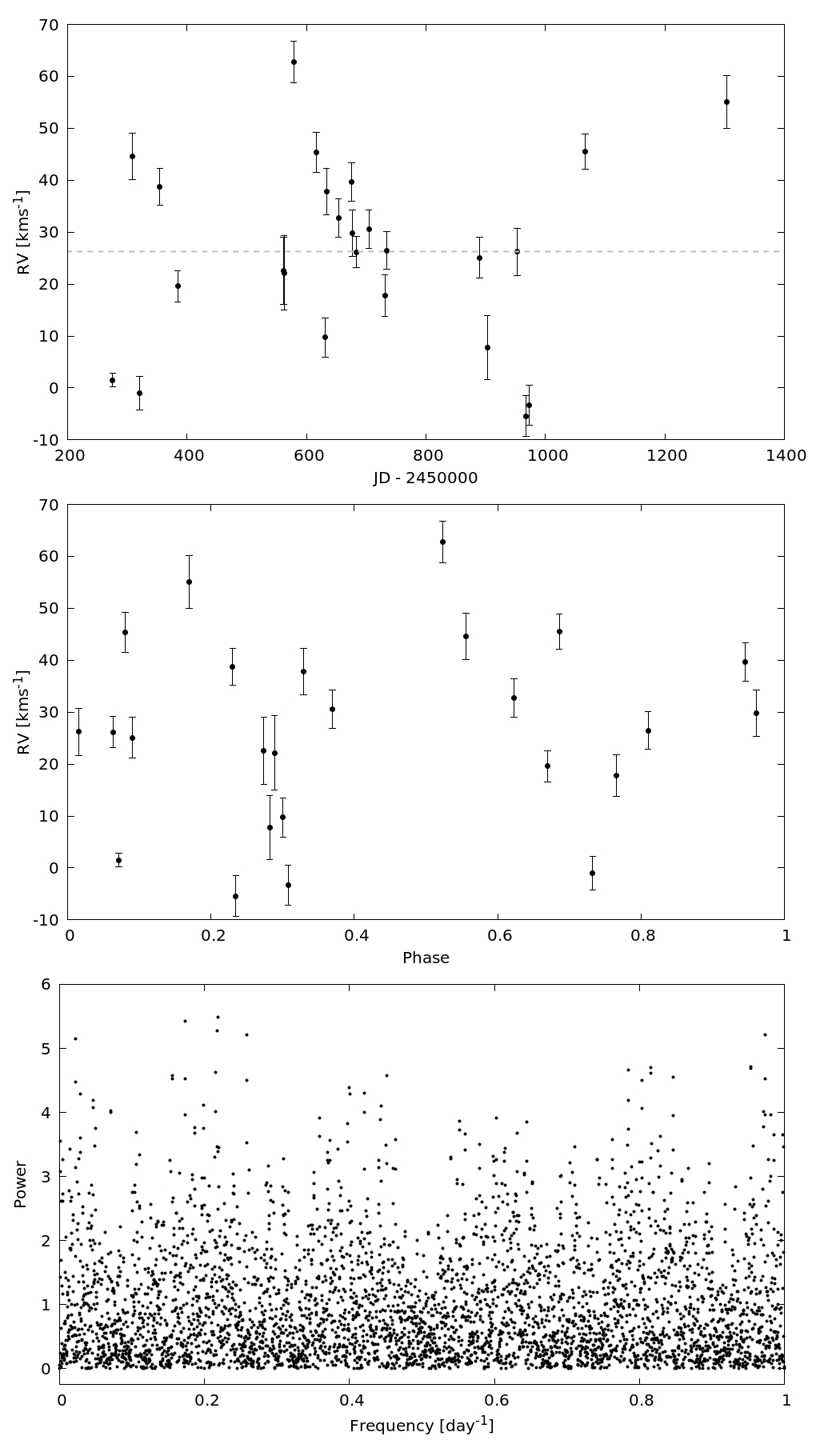}}
\caption{INT RV measurements. Top panel: RV versus time. Dashed grey line indicates the mean RV of 26.3~km~s\textsuperscript{-1}. Middle panel: RV versus 68-day brightness phase. Note the absence of any correlation between the RV variations and the 68-day brightness phase. Bottom panel: Lomb-Scargle periodogram. Note that the periodogram does not reveal any RV periodicity in the frequency range from 0 and 1~day\textsuperscript{-1}.}
\end{figure}

As discussed in subsection~3.1, some researchers have suggested that CS1, the star observed and presented in this paper, is actually not the central star of the PN (see Fig.~1). However, a useful result, derived from our RV analysis, is that the mean RV of the observed star, $\sim$26~km~s\textsuperscript{-1}, is in near agreement with the reported mean RV of the PN, 24.7~km~s\textsuperscript{-1} \citep{Schneider83}, as would be expected for a central star and its surrounding PN. We note that the RV of the much dimmer 19-magnitude CS2, an alternative candidate for the central star, is as yet unknown.

\section{SUGGESTED MODEL FOR THE CENTRAL STAR}

\subsection{Cataclysmic binary}

The cataclysmic binary model, suggested by \citet{Cuesta93}, provides a plausible explanation for the observed H$\alpha$ spectral line profiles as well as for the bipolar shape of the PN. However, a small separation between the stars, required for mass transfer to establish between a B8V and an ionising star, would correspond to an orbital period of less than 1.0 day. This would cause the RV of Balmer emission components to vary with $\pm\gtrsim$250$\cdot\sin(i)$~km~s\textsuperscript{-1}. Our set of AAT spectra did not reveal any noticeable spectral line profile or RV variations of Balmer emission lines in the 4.5-hour time-span which demonstrates that the cataclysmic binary model is highly unlikely. In addition, this model does not provide a plausible explanation for the 68-days brightness variation period.

\subsection{Be binary with precessing disc}

Our analysis of several potential models revealed that the observed spectrophotometric properties of CS1 are best reproduced with a close Be binary system with a misaligned precessing circumstellar disc. Other models such as Herbig Be stars, magnetically warped circumstellar discs, stellar pulsations etc. are discussed in \citet{Mocnik13}.

First of all, near edge-on rotating circumstellar discs around Be stars produce profiles of Balmer and singly ionised metal spectral lines similar to ones observed in our spectra. \citet{Hanuschik89} provided the relation between the full width at half maximum (FWHM) of the Balmer lines and the stellar rotational velocity. Our observed mean FWHM of H$\alpha$ (9.3~\AA) and H$\beta$ (5.7~\AA) emission profiles yield a stellar rotational velocity of $\sim$270~km~s\textsuperscript{-1}, which is in rough agreement with the measured rotational velocity of the central star ($200_{-25}^{+50}$~km~s\textsuperscript{-1}, see subsection 3.3.2).

Secondly, the observed spectral variability can be explained by various contributing factors, which have been reported for a significant fraction of Be stars, e.g. variable wind absorption, nonradial stellar pulsations, starspots, magnetically modulated circumstellar rotation, etc. (\citet{Porter03}, \citet{Rivinius13}). In a close Be binary system we might also expect to see the periodic spectral and RV variations arising from the orbital motion of the ionising star and consequent periodic heating of outer Be disc layers, similar to $\varphi$~Per, which is a binary between B0.5IVe and an O-type subdwarf (sdO6) \citep{Bozic95}. \citet{Poeckert81} reported that spectra of $\varphi$~Per also exhibit periodic interchanging blue- and red-dominated Balmer emission components, as indicated by Figs.~3 and 5 in the case of CS1. We do not see evidence that the spectral variations are periodic, however, our data were only sensitive to very short timescales ($<$1~day) and were sufficient to reject also the 68-day periodicity.

The expected orbital period of a close Be binary would be longer than in the case of a cataclysmic binary model. Having only two reasonable assumptions available, i.e. that the primary star had a mass higher than the B8V companion star in the pre-PN phase, $\sim$2.6~M$_\odot$ \citep{Belikov95}, and that the Roche lobe was filled during the red giant phase of the primary star, we can only derive rough theoretical lower and upper limits to the current orbital period to be 1 and 300 days, respectively. However, a similar close binary system of an early main sequence star (A5V) and an O-type subdwarf was found in the centre of the bipolar PN NGC~2346 with an orbital period 16 days \citep{Mendez81}. If a similar orbital period is applied to Sh2-71, this would comply with the absence of spectroscopic variations in hourly time-scales of AAT dataset and presence of variations in daily time-scales in the INT and LT datasets. A similar orbital period would also coincide with the tentatively identified eclipsing period of 17.2 days.

Furthermore, the mass transfer in the pre-PN evolution, required for the formation of a subdwarf, provides an explanation for the observed high metallicity in the spectra of the observed star, because the metal-rich material, recently dumped onto the B8V star in the pre-PN phase would not have had enough time to be fully mixed into the main-sequence star \citep{Smalley97}.

The photometric behaviour of the observed star can be explained by a variety of different contributing factors, such as starspots, stellar pulsations, reflection and/or irradiation if in a close binary system etc., but for the vast majority of cases, the effect would be insufficient or it would be correlated with spectral variations. On the other hand, a close Be binary system with a precessing Be disc has been identified as a possible plausible mechanism for the observed photometric properties and uncorrelated spectral variations of CS1. If the circumstellar Be disc is inclined with respect to orbital plane, the close binary ionising star can invoke tidal torques and cause the inclined disc to precess. The resulting periodic occultations of the companion star by the precessing disc can then be responsible for the observed 0.8-magnitude quasi-sinusoidal brightness variations. Furthermore, the detected variation of the reddening effect, $\Delta E(U-V)=0.38$, which coincides with the period and phase of the CS1's lightcurve, corresponds to an estimated absorption of 0.7--0.8~mag in \textit{V}-band and thus confirms that the extinction due to the precessing Be disc could be a realistic explanation for the observed photometric behaviour. The quasi-sinusoidal lightcurve with a period of 68 days would then correspond to a precession of the circumstellar disc, whereas the uncorrelated spectroscopic variations would result from the various intrinsic contributing factors listed above or, if periodic, possibly a result from orbital motion of the binary components.

The reported fraction of B8e among all B8 stars is 10 per cent, the vast majority of them being main sequence stars (Kogure et al. 1982), whereas the binary fraction of Be stars is $\sim$1/3 \citep{Porter03}. Combined fractions suggest that about 3 per cent of B8 stars are B8Ve binaries. From these, Be+sdO binaries in particular are supposed to exist in abundance, yet because they are not easily distinct, only three such systems are known with certainty \citep{Rivinius13}. This provides a small but still realistic probability that the central star candidate CS1 is a B8Ve+sdO binary.

The proposed precessing Be-disc model is highly tentative, however, it provides the explanation for the observed brightness variations and uncorrelated pronounced spectral variability. In addition, it also provides a potential collimation mechanism, required for the bipolar shape of Sh2-71. An extended daily-cadence sample of H$\alpha$ and H$\beta$ profiles would be required to verify any potential periodicity of the spectral variations and thus provide further evidence for the Be binary hypothesis.

\section{CONCLUSIONS}

Using photometric datasets from the central star candidate CS1 of the planetary nebula Sh2-71 from the Byrne Observatory at Sedgwick (BOS), the Faulkes Telescope South (FTS) and the Liverpool Telescope (LT), we tentatively identify the presence of periodic sharp brightness dips, possibly eclipses, with a period of $17.2\pm0.1$ days. We have also measured a periodic variable reddening effect with an amplitude $\Delta E(U-V)=0.38$, which coincides with the 68-days period and phase of the star's quasi-sinusoidal lightcurve.

Spectroscopic datasets from the LT, the Isaac Newton Telescope (INT) and the Anglo-Australian Telescope (AAT) confirmed pronounced variations in the shapes of the Balmer, helium and singly ionised metal lines. Our analysis revealed that these spectral variations are not correlated with the 68-days brightness variations and that the spectral variations occur on the time-scale of a few days. Comparing the observed with synthetic spectra yielded \textit{T}$_\rmn{eff}$ 11000--13000~K, $\log(g)$ $4.0\pm0.25$~cm~s\textsuperscript{-2}, $v_\rmn{rot}\cdot \sin(i)$ $200_{-25}^{+50}$~km~s\textsuperscript{-1} and indicated a high value of metallicity. The fitted parameters and the comparison with standard spectra classified the observed star as $\sim$B8V. The radial velocity (RV) was measured to vary with an amplitude of $\pm$40~km~s\textsuperscript{-1} around the mean value of $\sim$26~km~s\textsuperscript{-1}. The near agreement between the mean RV of the observed star and the reported mean RV of the nebula provides supporting evidence that the brighter candidate for the central star, CS1, indeed could be the true central star of Sh2-71. In addition, imaging with the INT revealed nebular features extending eastwards from CS1, which indicates that the observed star is likely to be physically linked with the planetary nebula.

We have revisited and rejected the cataclysmic model for the central star, proposed by \citet{Cuesta93}. The newly obtained spectrophotometric properties of the observed star were best reproduced by a close Be binary system with a precessing circumstellar disc. However, mainly due to the lack of spectra with a daily cadence, this newly suggested model should be regarded as tentative.

\section*{ACKNOWLEDGEMENTS}

We would like to thank the anonymous referee for their perceptive and detailed comments which led to improving this paper. We also thank Dr David Jones for his helpful suggestions in the RV analysis process. TM gratefully acknowledges the financial support of Slovene Human Resources Development and Scholarship Fund. This paper makes use of observations from the LCOGT network, the Liverpool Telescope, the Isaac Newton Telescope, the archived data obtained from the Anglo-Australian Telescope Data Archive, and the Isaac Newton Group Archive which is maintained as part of the CASU Astronomical Data Centre at the Institute of Astronomy, Cambridge. The Isaac Newton Telescope is operated on the island of La Palma by the Isaac Newton Group in the Spanish Observatorio del Roque de los Muchachos of the Instituto de Astrof\'{i}sica de Canarias. The Liverpool Telescope is operated on the island of La Palma by Liverpool John Moores University in the Spanish Observatorio del Roque de los Muchachos of the Instituto de Astrof\'{i}sica de Canarias with financial support from the UK Science and Technology Facilities Council.

\label{lastpage}



\section*{Supplementary material}

# The Central Star Candidate of the Planetary Nebula Sh2-71: Photometric and Spectroscopic Variability - Online Appendix

T. Močnik, M. Lloyd, D. Pollacco and R. A. Street

### Abstract

This online appendix lists the differential magnitudes and radial velocity measurements of the central star candidate CS1 and expands the spectroscopic analysis, introduced in the main article. Additional data, explanation and figures are focused in four sections: Differential Photometry, Radial Velocities, Synthetic Spectra and Standard Spectra.

## 1   DIFFERENTIAL PHOTOMETRY

Differential magnitudes for the central star candidate CS1 have been measured using several comparison stars. However, the most consistent photometric results and the highest signal to noise ratio was obtained using comparison star number 2 for BOS and comparison star number 1 for FTS and LT. Tables 1.1-1.5 list the final differential magnitudes for star CS1.

## 2   RADIAL VELOCITIES

Radial velocities (RVs) of the central star candidate CS1 have been measured as relative RVs using cross-correlation technique and the first spectrum as a template spectrum in each of the two datasets, INT and AAT. Relative RVs were then scaled into absolute values by measuring the wavelength of Mg II 4481 Å line in every spectrum in both datasets. Tables 2.1 and 2.2 list the absolute heliocentric RVs from the INT and AAT datasets, respectively.

## 3   SYNTHETIC SPECTRA

Normalized spectra of the central star candidate CS1 have been visually compared to synthetic spectra with dispersion 1 Å/pixel, computed by Munari et al. (2005) with the Kurutz code. Comparison between the measured spectrum and synthetic spectra of various rotational velocities is shown in Fig. 3.1, comparison to synthetic spectra with various surface gravities in Fig. 3.2 and various effective temperatures in Figs. 3.3 and 3.4. Table 3.1 demonstrates the influence of the effective temperature on the ratio between He I (4471 Å) and Mg II (4481 Å) lines. The effective temperature was determined to vary within a 2000 K range primarily due to the measured variable ratio between He I and Mg II lines (see Table 3.2). The metallicity was determined to be larger than the largest available value of metallicity in the library of synthetic spectra, [M/H] > 0.5. The comparison between measured and synthetic spectra with the final selection of atmospheric parameters in the entire wavelength range is shown in Fig. 3.5. The atmospheric parameters, used for the computation of the best-fit spectra, are listed in Table 3.3 and correspond to B9V–B7V (Lang, 1992).

## 4   STANDARD SPECTRA

The measured spectra of the central star candidate CS1 have been compared to a number of standard stellar spectra of various spectral types, provided by Bagnulo et al. (2003). The comparison was used to verify the ∼B8V spectral type of CS1, determined by comparisons with synthetic spectra. Fig. 4.1 shows the comparison between the measured INT spectrum of the central star candidate CS1 and the selected sample of standard spectra of spectral types A7V, B8V and B6V.

Table 1.1: Log of differential magnitudes for the central star candidate CS1 using the BOS, filter $B$ and comparison star number 2.

| JD - 2450000 | $\Delta B$ | JD - 2450000 | $\Delta B$ |
|---|---|---|---|
| | mag(2) - mag(CS1) | | mag(2) - mag(CS1) |
| 5739.816 | 0.8217±0.0066 | 5751.812 | 0.9764±0.0063 |
| 5743.837 | 0.8940±0.0058 | 5761.734 | 1.2197±0.0065 |
| 5743.902 | 0.9102±0.0061 | 5762.762 | 1.1494±0.0061 |
| 5743.938 | 0.9067±0.0064 | 5764.733 | 1.1661±0.0058 |
| 5748.897 | 0.8611±0.0064 | 5764.741 | 1.1679±0.0058 |
| 5748.904 | 0.8462±0.0066 | 5765.764 | 1.2898±0.0057 |
| 5749.768 | 0.9490±0.0065 | 5768.712 | 1.4281±0.0062 |
| 5749.842 | 0.9660±0.0062 | 5769.716 | 1.3524±0.0058 |
| 5749.856 | 0.9776±0.0061 | 5770.721 | 1.4605±0.0064 |
| 5751.798 | 0.9816±0.0069 | 5771.743 | 1.4890±0.0060 |
| 5751.805 | 0.9802±0.0064 | | |

Table 1.2: Log of differential magnitudes for CS1 using the FTS, filter $V$ and comparison star number 1.

| JD - 2450000 | $\Delta V$ | JD - 2450000 | $\Delta V$ |
|---|---|---|---|
| | mag(1) - mag(CS1) | | mag(1) - mag(CS1) |
| 5838.890 | -2.2670±0.0095 | 5997.249 | -1.9800±0.0085 |
| 5842.947 | -2.0290±0.0095 | 5998.292 | -2.1220±0.0095 |
| 5843.896 | -1.9250±0.0250 | 6010.305 | -2.5360±0.0114 |
| 5844.897 | -1.8890±0.0085 | | |

Table 1.3: Log of differential magnitudes for CS1 using the FTS, filter $B$ and comparison star number 1.

| JD - 2450000 | $\Delta B$ | JD - 2450000 | $\Delta B$ |
|---|---|---|---|
| | mag(1) - mag(CS1) | | mag(1) - mag(CS1) |
| 5838.889 | -2.3440±0.0155 | 5997.248 | -2.0730±0.0168 |
| 5842.946 | -2.0760±0.0168 | 5998.292 | -2.1870±0.0228 |
| 5843.896 | -1.9720±0.0323 | 6010.304 | -2.5940±0.0226 |
| 5844.897 | -1.9510±0.0139 | | |

Table 1.4: Log of differential magnitudes for CS1 using the LT, filter $V$ and comparison star number 1.

| JD - 2450000 | $\Delta V$ | JD - 2450000 | $\Delta V$ |
|---|---|---|---|
| | mag(1) - mag(CS1) | | mag(1) - mag(CS1) |
| 6018.721 | -2.5310±0.0116 | 6092.672 | -2.5465±0.0113 |
| 6046.594 | -2.0740±0.0095 | 6093.683 | -2.5467±0.0118 |
| 6049.576 | -1.8556±0.0080 | 6095.463 | -2.5685±0.0151 |
| 6052.568 | -2.1912±0.0109 | 6095.562 | -2.5351±0.0113 |
| 6056.665 | -2.3214±0.0103 | 6103.648 | -2.3478±0.0115 |
| 6056.668 | -2.3117±0.0102 | 6106.541 | -2.4902±0.0144 |
| 6060.608 | -2.1531±0.0268 | 6108.567 | -2.5055±0.0138 |
| 6074.515 | -2.3415±0.0118 | 6109.574 | -2.4398±0.0121 |
| 6076.508 | -2.4097±0.0113 | 6110.522 | -2.2982±0.0111 |
| 6076.528 | -2.3913±0.0111 | 6110.525 | -2.3294±0.0120 |
| 6079.493 | -2.4715±0.0127 | 6114.524 | -2.0552±0.0093 |
| 6082.700 | -2.5325±0.0111 | 6116.584 | -1.9536±0.0091 |
| 6083.600 | -2.5409±0.0113 | 6117.558 | -1.9007±0.0092 |
| 6084.640 | -2.5048±0.0111 | 6121.565 | -1.8546±0.0091 |
| 6087.645 | -2.5043±0.0111 | 6124.565 | -1.9821±0.0117 |
| 6090.694 | -2.5145±0.0113 | 6128.497 | -1.7330±0.0100 |
| 6090.697 | -2.5012±0.0110 | 6128.500 | -1.7287±0.0093 |
| 6091.668 | -2.5048±0.0105 | 6129.575 | -1.8577±0.0127 |

Table 1.5: Log of differential magnitudes for CS1 using the LT, filter $U$ and comparison star number 1.

| JD - 2450000 | $\Delta U$ | JD - 2450000 | $\Delta U$ |
| | mag(1) - mag(CS1) | | mag(1) - mag(CS1) |
| --- | --- | --- | --- |
| 6018.722 | -2.9742±0.0277 | 6091.668 | -2.8498±0.0256 |
| 6046.595 | -2.3111±0.0267 | 6092.673 | -2.9139±0.0250 |
| 6049.576 | -1.9648±0.0186 | 6093.683 | -2.9632±0.0293 |
| 6052.569 | -2.2126±0.0228 | 6095.562 | -3.0038±0.0279 |
| 6056.666 | -2.5361±0.0212 | 6103.649 | -2.7321±0.0279 |
| 6056.668 | -2.4795±0.0203 | 6106.541 | -2.7709±0.0301 |
| 6060.612 | -2.4066±0.0443 | 6108.568 | -2.8010±0.0278 |
| 6074.516 | -2.6633±0.0266 | 6109.574 | -2.6252±0.0246 |
| 6076.509 | -2.7534±0.0274 | 6110.523 | -2.4069±0.0235 |
| 6076.529 | -2.7148±0.0252 | 6110.525 | -2.4842±0.0229 |
| 6079.494 | -2.8251±0.0320 | 6114.525 | -2.2627±0.0212 |
| 6082.700 | -2.8596±0.0273 | 6116.585 | -2.1451±0.0184 |
| 6083.600 | -2.9087±0.0258 | 6117.559 | -2.0160±0.0175 |
| 6084.640 | -2.8747±0.0253 | 6121.566 | -2.0329±0.0174 |
| 6087.645 | -2.8669±0.0239 | 6128.498 | -2.0713±0.0190 |
| 6090.695 | -2.9846±0.0264 | 6128.500 | -2.0572±0.0192 |
| 6090.697 | -2.9328±0.0257 | 6129.575 | -2.1450±0.0258 |

Table 2.1: Log of heliocentric RVs for the central star candidate CS1 using the INT. The first error bar corresponds to the error of absolute RV scaling by Mg ɪɪ line, the second error bar is the cross-correlation error.

| JD - 2450000 | RV (km s$^{-1}$) | JD - 2450000 | RV (km s$^{-1}$) |
| --- | --- | --- | --- |
| 275.4937 | 1.48±1.31 | 676.5633 | 29.78±1.31±3.16 |
| 308.4805 | 44.58±1.31±3.18 | 683.5339 | 26.17±1.31±1.65 |
| 320.5166 | -1.03±1.31±1.90 | 704.3787 | 30.56±1.31±2.39 |
| 354.4195 | 38.79±1.31±2.25 | 731.3648 | 17.76±1.31±2.66 |
| 384.3285 | 19.58±1.31±1.66 | 734.3913 | 26.49±1.31±2.34 |
| 561.6797 | 22.59±1.31±5.20 | 889.6922 | 25.11±1.31±2.62 |
| 562.6892 | 22.15±1.31±5.88 | 902.7299 | 7.80±1.31±4.82 |
| 578.6710 | 62.78±1.31±2.66 | 952.6966 | 26.23±1.31±3.23 |
| 616.6297 | 45.33±1.31±2.52 | 967.6333 | -5.43±1.31±2.59 |
| 631.6013 | 9.71±1.31±2.43 | 972.5833 | -3.34±1.31±2.51 |
| 633.5566 | 37.80±1.31±3.16 | 1066.4821 | 45.52±1.31±2.14 |
| 653.5404 | 32.79±1.31±2.39 | 1303.7116 | 55.06±1.31±3.81 |
| 675.5400 | 39.72±1.31±2.40 | | |

Table 2.2: Log of heliocentric RVs for CS1 using the AAT. The first error bar corresponds to the error of absolute RV scaling by Mg ɪɪ line, the second error bar is the cross-correlation error.

| JD - 2450000 | RV (km s$^{-1}$) | JD - 2450000 | RV (km s$^{-1}$) |
| --- | --- | --- | --- |
| 1686.14051 | 12.09±1.01 | 1686.26094 | 11.76±1.01±0.22 |
| 1686.16222 | 11.46±1.01±0.19 | 1686.28265 | 11.55±1.01±0.21 |
| 1686.17858 | 12.13±1.01±0.20 | 1686.30438 | 11.28±1.01±0.21 |
| 1686.21750 | 11.32±1.01±0.20 | 1686.32536 | 12.09±1.01±0.30 |
| 1686.23923 | 11.14±1.01±0.21 | | |

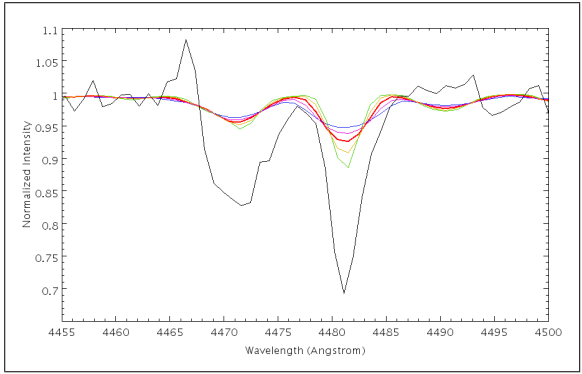

Figure 3.1: Determination of the rotational velocity. The first INT spectrum, obtained on 1996 July 10 (black), is compared to synthetic spectra with stellar rotational velocities ranging from 100 km s⁻¹ (green) to 300 km s⁻¹ (blue). The basal widths of the absorption lines were best reproduced by a rotational velocity of 200 km s⁻¹ (red). Other atmospheric parameters are the same for all synthetic spectra: effective temperature 11000 K, log(g) 4.0 cm s⁻² and solar metallicity. The two plotted spectral features are He I (4471 Å) and Mg II (4481 Å).

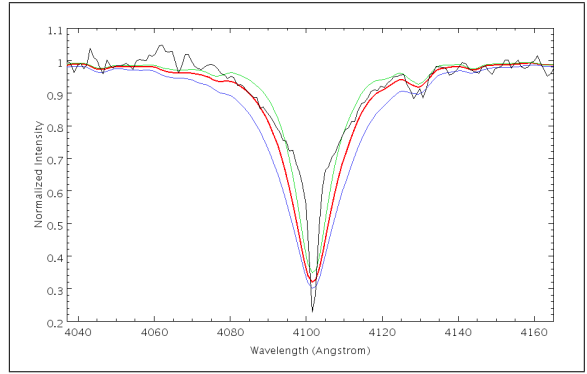

Figure 3.2: Determination of the surface gravity. The Hδ profile is fitted by synthetic spectra with 3.5 (green), 4.0 (red) and 4.5 cm s⁻² (blue) logarithmic stellar surface gravitational acceleration. The best fit of the Balmer wings was achieved with log(g) = 4.0 cm s⁻² (red). The central parts of the Balmer line profiles were affected by the emission components and were excluded from the fitting process.

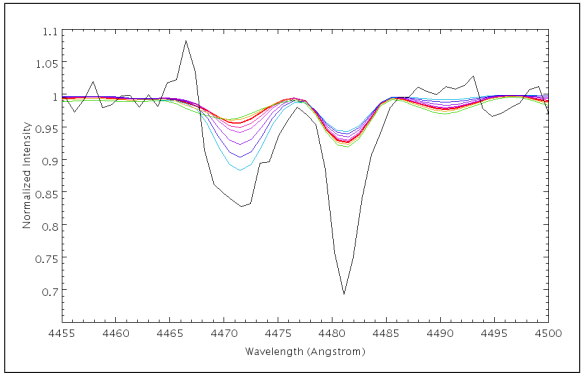

Figure 3.3: The effect of the effective temperature on the He I/Mg II ratio for the temperatures given in Table 3.1, from 10000 K (green) to 15000 K (blue). A higher effective temperature increases the strength of the He I line (4471 Å), whereas the strength of the Mg II line (4481 Å) is slightly reduced. The measured He I/Mg II ratio from the first INT spectrum obtained on 1996 July 10 is best reproduced by the synthetic spectrum with effective temperature 11000 K (red line). Note how the He I line becomes affected by the Ti II line (4468.5 Å) at lower effective temperatures.

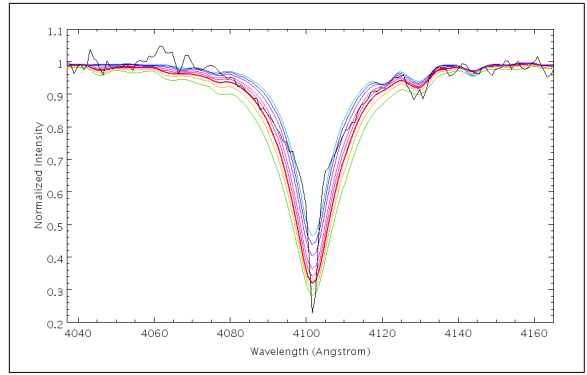

Figure 3.4: The impact of the effective temperature on the Balmer lines, from 10000 K (green) to 15000 K (blue). The absorption wings of the Hδ profile are well fitted by the 11000 K synthetic spectrum (red). This confirms the choice of 11000 K, which was made based on the He I/Mg II ratio (see Table 3.2 and Fig. 3.3). As for the gravity fitting process, the central region of the Balmer profiles have not been fitted, because of the presence of the emission and self-absorption components.

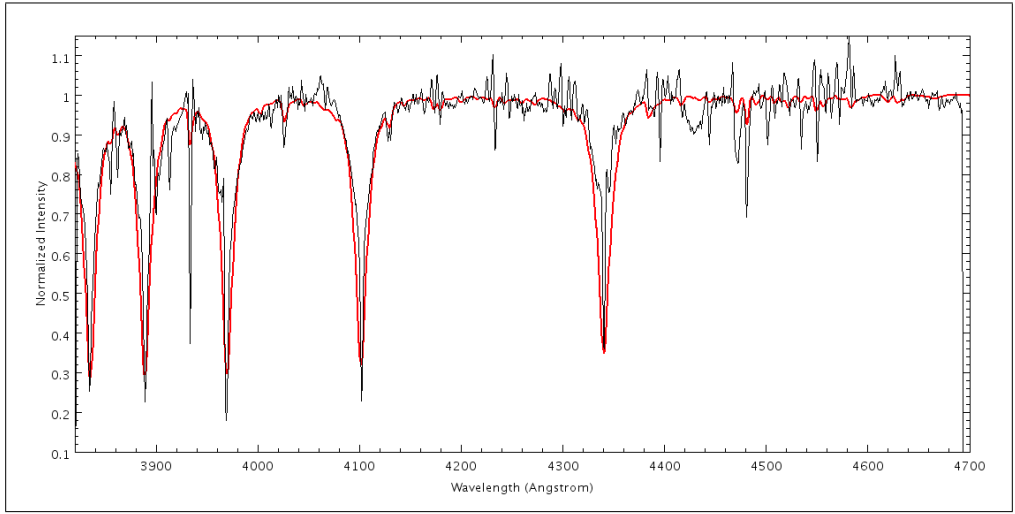

Figure 3.5: The best fit synthetic spectrum (red) is shown against the measured INT spectrum from 1996 July 10 (black). The atmospheric parameters of the synthetic spectrum are $T_{\text{eff}}$=11000 K, $\log(g)$=4.0 cm s$^{-2}$, $v_{\text{rot}}$=200 km s$^{-1}$ and [M/H]=0. The discrepancies between the measured and synthetic metal line strengths are a result of the high metallicity of the observed star.

Table 3.1: He I (4471 Å) to Mg II (4481 Å) ratio dependence on the effective temperature. The ratios are calculated from synthetic spectra with parameters $\log(g)$ = 4.0 cm s$^{-2}$, [M/H]=0 and $v_{\text{rot}}$ = 200 km s$^{-1}$.

| $T_{\text{eff}}$ [K] | He I/Mg II | $T_{\text{eff}}$ [K] | He I/Mg II |
|---|---|---|---|
| 10000 | 0.43 | 12000 | 0.82 |
| 10500 | 0.45 | 13000 | 1.07 |
| 11000 | 0.53 | 14000 | 1.66 |
| 11500 | 0.65 | 15000 | 2.70 |

Table 3.2: Estimated effective temperature. The measured ratio between He I (4471 Å) and Mg II (4481 Å) and the corresponding effective temperature (see Table 3.1) is provided for various INT spectra. The INT spectrum, obtained on 1996 October 27, exhibits the smallest line ratio, while the largest ratio can be seen in the spectrum, taken on 1998 March 16.

| observation date | He I/Mg II | $T_{\text{eff}}$ [K] |
|---|---|---|
| 1996 Jul 10 | 0.55 | 11000 |
| 1996 Oct 27 (min) | 0.51 | 11000 |
| 1998 Mar 16 (max) | 1.12 | 13000 |
| average | 0.80 | 12000 |

Table 3.3: The best fit atmospheric parameters of CS1. The effective temperature, stellar surface gravity and projected rotational velocity have been fitted with the synthetic spectra. Metallicity could not be fitted, because it exceeded the values, provided in the library of synthetic spectra.

| $T_{\text{eff}}$ [K] | 11000–13000 |
|---|---|
| $\log(g)$ [cm s$^{-2}$] | $4.0 \pm 0.25$ |
| $v_{\text{rot}} \cdot \sin(i)$ [km s$^{-1}$] | $200^{+50}_{-25}$ |
| [M/H] | $\gg 0$ |

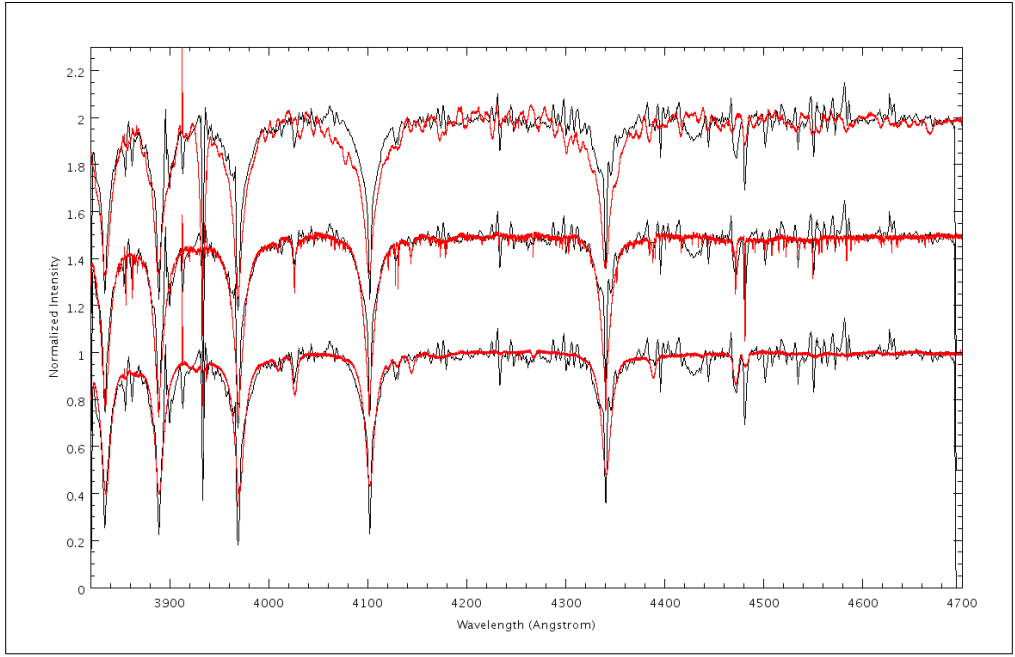

Figure 4.1: Comparison between our measured INT spectrum of CS1 (black) and standard spectra A7V (top), B8V (middle) and B6V (bottom) shown in red. A7V: Note the absence of Si II lines (3856 and 3863 Å), He I (4026 and 4471 Å) and presence of Mn I (4033 Å), Ca I (4227 Å) and Ti II (4468 Å). The Balmer profiles exhibit much stronger gravitational broadening than in our measured spectra. B8V: Si II and He I lines are present, whereas Ti II (4468 Å) is not because of the higher temperatures of B8 stars. The measured depth ratio between He I (4471 Å) and Mg II line (4481 Å) is perfectly matched. Also note the comparable amount of gravitational broadening in the Balmer absorption lines. The narrowness of the other lines may be disregarded because of the smaller projected rotational velocity of the B8V standard star. B6V: The 2000 K higher effective temperature, compared to B8, results in highly pronounced He I lines (4026 and 4471 Å), whereas the lines of the heavier elements are no longer present. The larger radius and the corresponding smaller gravitational broadening of a B6V star mean that the absorption wings of the Balmer profiles are not as well matched as for the B8V star. The most noticeable discrepancy, the large ratio between the He I and Mg II line, suggests that the spectral type of the observed star has to be later than B6.




\begin{thebibliography}{99}
\bibitem[\protect\citeauthoryear{Bagnulo et al.}{2003}]{Bagnulo03} Bagnulo S., Jehin E., Ledoux C., Cabanac R., Melo C., Gilmozzi R., ESO Paranal Science Operations Team, 2003, The Messenger, 114, 10
\bibitem[\protect\citeauthoryear{Barnsley et al.}{2012}]{Barnsley12} Barnsley R.M., Smith R.J., Steele I.A., 2012, Astronomische Nachrichten, 333, 101
\bibitem[\protect\citeauthoryear{Belikov}{1995}]{Belikov95} Belikov A.N., 1995, VizieR Online Data Catalog, 5085, 0
\bibitem[\protect\citeauthoryear{Bozic et al.}{1995}]{Bozic95} Bo\v{z}i\'{c} H., Harmanec P., Horn J., Koubsky P., Scholz G., McDavid D., Hubert A.M., Hubert H., 1995, A\&A, 304, 235
\bibitem[\protect\citeauthoryear{Cuesta \& Phillips}{1993}]{Cuesta93} Cuesta L., Phillips J.P., 1993, A\&A, 270, 379
\bibitem[\protect\citeauthoryear{De Marco}{2009}]{DeMarco09} De Marco O., 2009, PASP, 121, 316
\bibitem[\protect\citeauthoryear{De Marco et al.}{2008}]{DeMarco08} De Marco O., Hillwig T.C., Smith A.J., 2008, AJ, 136, 323
\bibitem[\protect\citeauthoryear{Frew \& Parker}{2007}]{Frew07} Frew D.J., Parker Q.A., 2007, in Asymmetrical Planetary Nebulae IV Do post-common envelope objects form a distinct subset of PNe?
\bibitem[\protect\citeauthoryear{Hanuschik}{1989}]{Hanuschik89} Hanuschik R.W., 1989, Ap\&SS, 161, 61
\bibitem[\protect\citeauthoryear{Jurcsik}{1993}]{Jurcsik93} Jurcsik J., 1993, in Weinberger R., Acker A., eds, Planetary Nebulae Vol. 155 of IAU Symposium, On the Photometric Behaviour of the Central Star of the Planetary Nebula Sh2-71 p. 399
\bibitem[\protect\citeauthoryear{Kohoutek}{1979}]{Kohoutek79} Kohoutek L., 1979, Information Bulletin on Variable Stars, 1672, 1
\bibitem[\protect\citeauthoryear{Kohoutek}{2001}]{Kohoutek01} Kohoutek L., 2001, VizieR Online Data Catalog, 4024, 0
\bibitem[\protect\citeauthoryear{Mendez \& Niemela}{1981}]{Mendez81} Mendez R.H., Niemela V.S., 1981, ApJ, 250, 240
\bibitem[\protect\citeauthoryear{Mikul{\'a}{\v s}ek et al.}{2007}]{Mikulasek07} Mikul{\'a}{\v s}ek Z., Skopal A., Zejda M., Pejcha O., Kohoutek L., Motl D., Vittone A.A., Errico L., 2007, in Okazaki A.T., Owocki S.P., Stefl S., eds, Active OB-Stars: Laboratories for Stellar and Circumstellar Physics Vol. 361 of Astronomical Society of the Pacific Conference Series, Light Variations of the Anomalous Central Star of Planetary Nebula Sh 2-71. p. 469
\bibitem[\protect\citeauthoryear{Minkowski}{1946}]{Minkowski46} Minkowski R., 1946, PASP, 58, 305
\bibitem[\protect\citeauthoryear{Mo\v{c}nik}{2013}]{Mocnik13} Mo\v{c}nik T., 2013, Master's thesis, The University of Manchester. Available at:\\https://www.escholar.manchester.ac.uk/uk-ac-man-scw:215279
\bibitem[\protect\citeauthoryear{Munari et al.}{2005}]{Munari05} Munari U., Sordo R., Castelli F., Zwitter T., 2005, A\&A, 442, 1127
\bibitem[\protect\citeauthoryear{Poeckert}{1981}]{Poeckert81} Poeckert R., 1981, PASP, 93, 297
\bibitem[\protect\citeauthoryear{Porter \& Rivinius}{2003}]{Porter03} Porter J.M., Rivinius T., 2003, PASP, 115, 1153
\bibitem[\protect\citeauthoryear{Rivinius et al.}{2013}]{Rivinius13} Rivinius T., Carciofi A.C., Martayan C., 2013, A\&A Rev., 21, 69
\bibitem[\protect\citeauthoryear{Schneider et al.}{1983}]{Schneider83} Schneider S.E., Terzian Y., Purgathofer A., Perinotto M., 1983, ApJS, 52, 399
\bibitem[\protect\citeauthoryear{Sharpless}{1959}]{Sharpless59} Sharpless S., 1959, ApJS, 4, 257
\bibitem[\protect\citeauthoryear{Smalley}{1997}]{Smalley97} Smalley B., 1997, The Observatory, 117, 338
\end{thebibliography}
\end{document}